\documentclass[a4paper]{PoS}
\usepackage{graphics}
\usepackage{graphicx}

\title{Survival analysis of the optical brightness of GRB host galaxies}

\ShortTitle{Survival analysis of the optical brightness of GRB host galaxies}

\author{\speaker{Istv\'{a}n R\'{a}cz}\\
        Konkoly Observatory of the Hungarian Academy of Science, H-1121 Budapest, Hungary\\
        E-mail: \email{racz.istvan@csfk.mta.hu}}

\author{Lajos G. Bal\'{a}zs\\
        Konkoly Observatory of the Hungarian Academy of Science, H-1121 Budapest, Hungary\\
        E\"{o}tv\"{o}s University, H-1117 Budapest, Hungary}

\author{Zsolt Bagoly\\
		E\"{o}tv\"{o}s University, H-1117 Budapest, Hungary}

\author{L. Viktor T\'{o}th\\
		E\"{o}tv\"{o}s University, H-1117 Budapest, Hungary}

\abstract{We studied the unbiased optical brightness distribution which was calculated from the survival analysis of host galaxies and its relationship with the Swift GRB data of the host galaxies observed by the Keck telescopes. Based on the sample obtained from merging the Swift GRB table and the Keck optical data we also studied the dependence of this distribution on the data of the GRBs. Finally, we compared the HGs distribution with standard galaxies distribution which is in the DEEP2 galaxies catalog.}

\FullConference{Swift: 10 Years of Discovery,\\
		2-5 December 2014\\
		La Sapienza University, Rome, Italy }

\begin{document}

\section{Introduction}

Most LGRBs are identified with massive stellar explosions, frequently in star-forming galaxies. Many host galaxies (HGs) are observed and it is an open question yet whether the luminosity distribution of HGs, characterized by GRB events, is identical with those of the galaxies in general. Unfortunately, many HGs are fainter than the detection limit of optical observation. 

Nevertheless, we can give an upper limit of their luminosity \cite{savaglio}. Here the survival analysis can help to answer this question giving an unbiased estimate of the luminosity distribution.

\section{Mathematical summary}
\begin{figure}[!h]
\includegraphics[width=.51\textwidth]{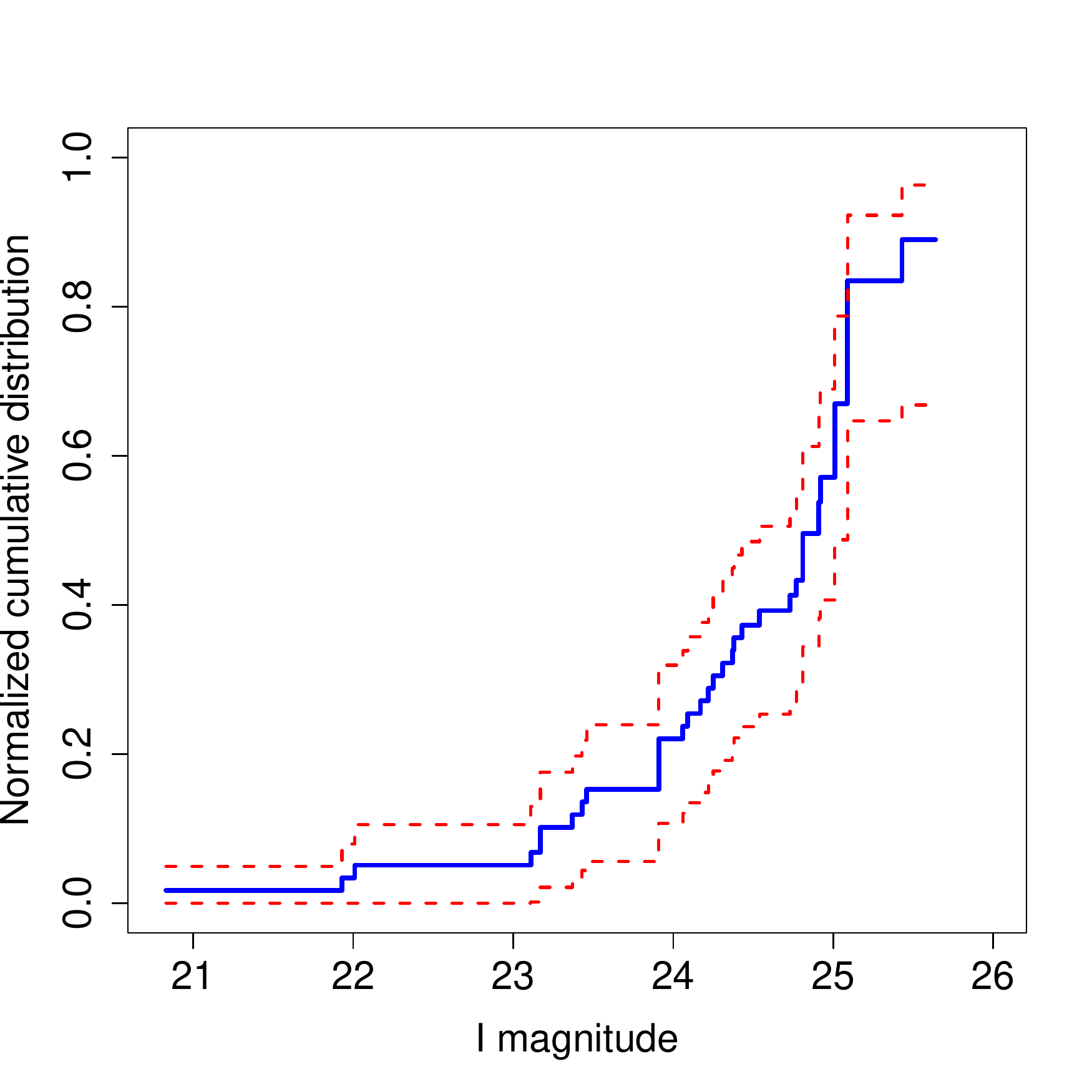}
\includegraphics[width=.51\textwidth]{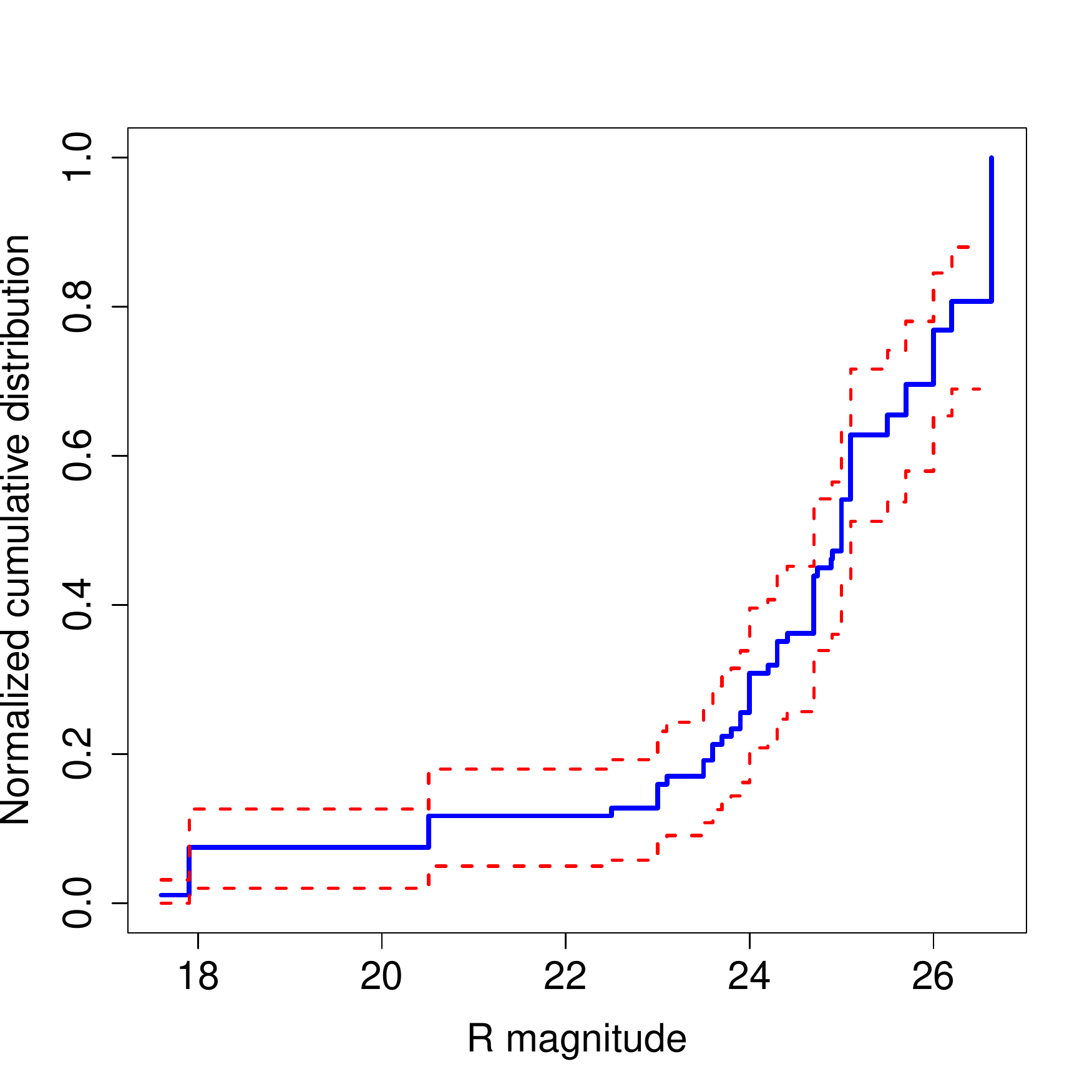}
\caption{Distributions calculated from the survival function with 95\% confidence bounds in the I (left) and R (right) colors.}
\label{fig1}
\end{figure}

The Survival analysis is a statistical procedure where some data in the sample of the measurements are censored (have only an upper/lower bound). Such data sets are also common in astrophysical problems. The detection limitations in the form of upper or lower limits are often called left or right censoring: the luminosity expressed in magnitude is a right censored event. The output of this procedure is an unbiased estimate of the survival function: 
\[S(t) =  Prob(T > t)\]
where t is the measured value (magnitude in our cases) of the T random variable. The distribution function (Figure \ref{fig1}) which also called as lifetime distribution function is: \[F(t) = 1-S(t)\] For the more astronomical context see Feigelson and Nelson \cite{fei}.

The basic Cox regression proportional hazards (Cox PH) model fits survival data depending on z co-variate to a hazard function (h(t|z)) in the form of 
\[\frac{\dot{F}(t)}{S(t)}=h(t|z)=h_{0}\cdot e^{(\beta'z)}\]
where $\beta$ is an N-element unknown vector and h$_0$(t) is called the baseline hazard, which is non-parametric. Primary interest lies in estimating the parameter using the partial likelihood function:
\[ L(\beta)=\prod^{N}_{i=0} \frac{e^{\beta'z_{i}}}{\sum_{j\in R(t_{i})}e^{\beta'z_{j}}}\]
where R(t$_\mathrm{i}$) is the risk set at t$_\mathrm{i}$ \cite{isobe}.
If the estimate of the baseline survival function S$_\mathrm{0}$(t) is provided, then the estimate of the survival function for an individual with co-variates z$_\mathrm{k}$ may be obtained via:
\[ S(t|z_k)=[S_{0}(t)]^{exp(\beta'z_{k})}\]

\section{Observational data}
For searching host galaxies more than 160 GRB positions were observed in the Preliminary Keck GRB Host Project. Most observations were carried out between 2005 and 2010 under a series of proposals focusing on host discovery and basic characterization (via the observers-frame optical color), and putting redshift constraints, in particular to rule out a large high-z fraction which was suggested in some early works \cite{perley}. We used the Preliminary Keck GRB Host Project Imaging Catalog with 164 host galaxies. The photometric systems were Johnson's UBV and Cousins' RI. 

The DEEP2 survey used the Keck II telescope to study the distant Universe. First the LRIS spectrograph was used in Phase 1 and a sample of 1000 galaxies studied within a limit of I=24.5 magnitude. Then in Phase 2 of the DEEP project was using the DEIMOS spectrograph to obtain spectra of 50,000 faint galaxies with redshift of z>0.7 and a limiting apparent magnitude of R$_\mathrm{AB}$=24.1 mag. This spectrograph can observe in three bands (BRI). The survey covers an area of 2.8deg$^\mathrm{2}$ divided into four separate fields observed \cite{newman}. We assumed that all measured data are reliable namely not censoring data so we could use over than 49,000 galaxies in I band and about 48,000 galaxies in R band.

\section{Survival analysis data} 
We used the R and I magnitudes of Preliminary Keck GRB Host Project representing a sufficiently large homogeneous sample for survival analysis in R programming language with 'survival' package \cite{surv}. We have 58 R and 38 I color observation having 14 and 7 lower limits (censored) for the magnitudes, respectively. We calculated both (R and I) survival functions from these magnitudes.

We used the DEEP2 galaxy catalog for making the comparison to the results of the survival analysis. The compared range was between about 21 and 24 magnitude in both colors because here are enough many data in both database.

Testing the luminosity distributions no significant difference was found between HGs with PL or CPL LGRB spectra. The Cox regression was used to determine whether the properties of GRBs (SWIFT BAT T90, Fluence, 1-sec Peak Photon Flux, Photon Index, XRT 11 Hour Flux, XRT 24 Hour Flux and redshift) depend on the HG distribution.

\section{Results}

\begin{figure}[!h]
\includegraphics[width=.51\textwidth]{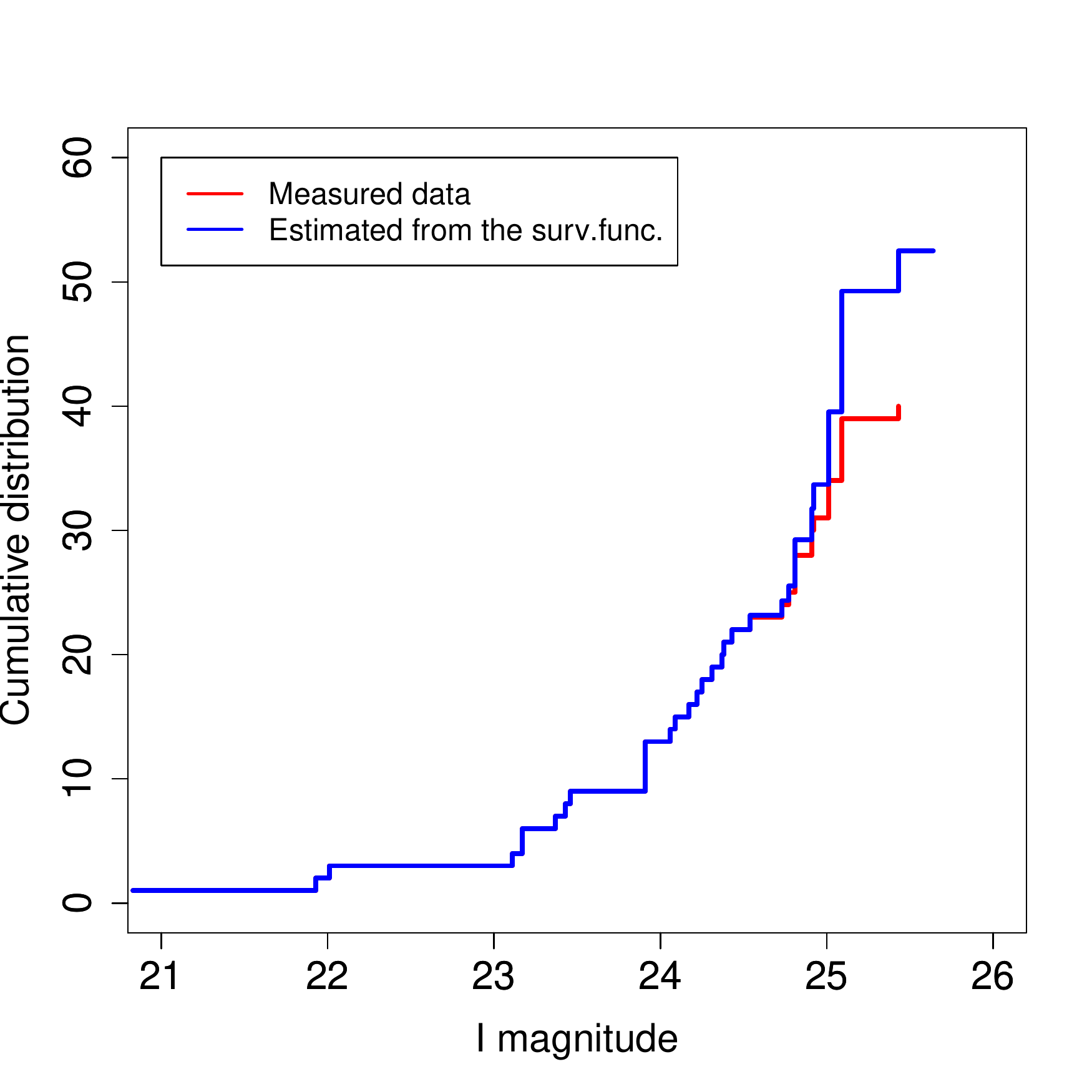}
\includegraphics[width=.51\textwidth]{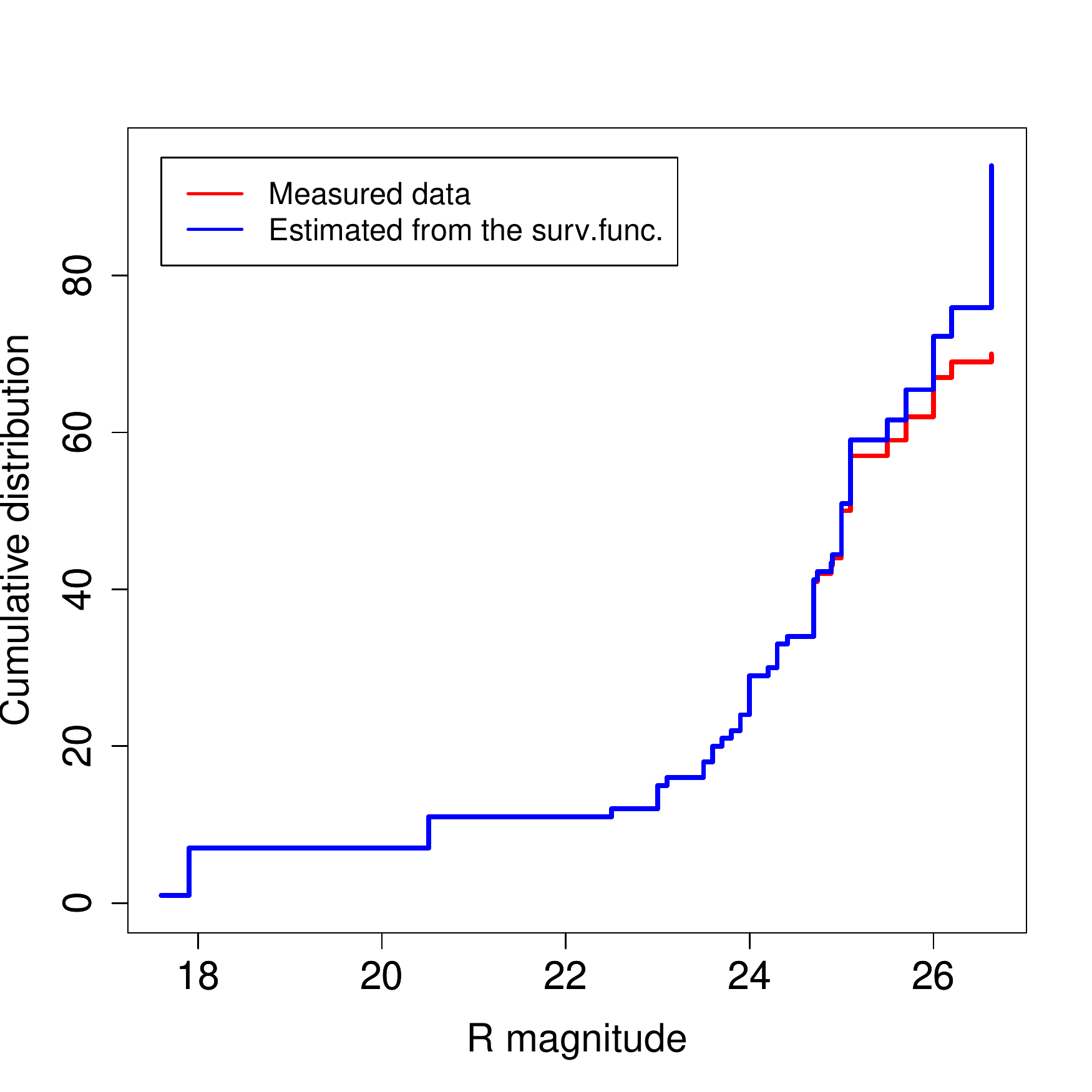}
\caption{The comparison of the distribution function obtained from the survival function (blue) with the cumulative measured data (red) of I color (left) and R color (right).}
\label{fig2}
\end{figure}

Figure \ref{fig2} shows the difference between the cumulative distribution of the measured data and survival function in both color. It seems that the fainter are the objects more different are the two functions because the most censored data are among fainter points.

\begin{figure}[h]
\includegraphics[width=.51\textwidth]{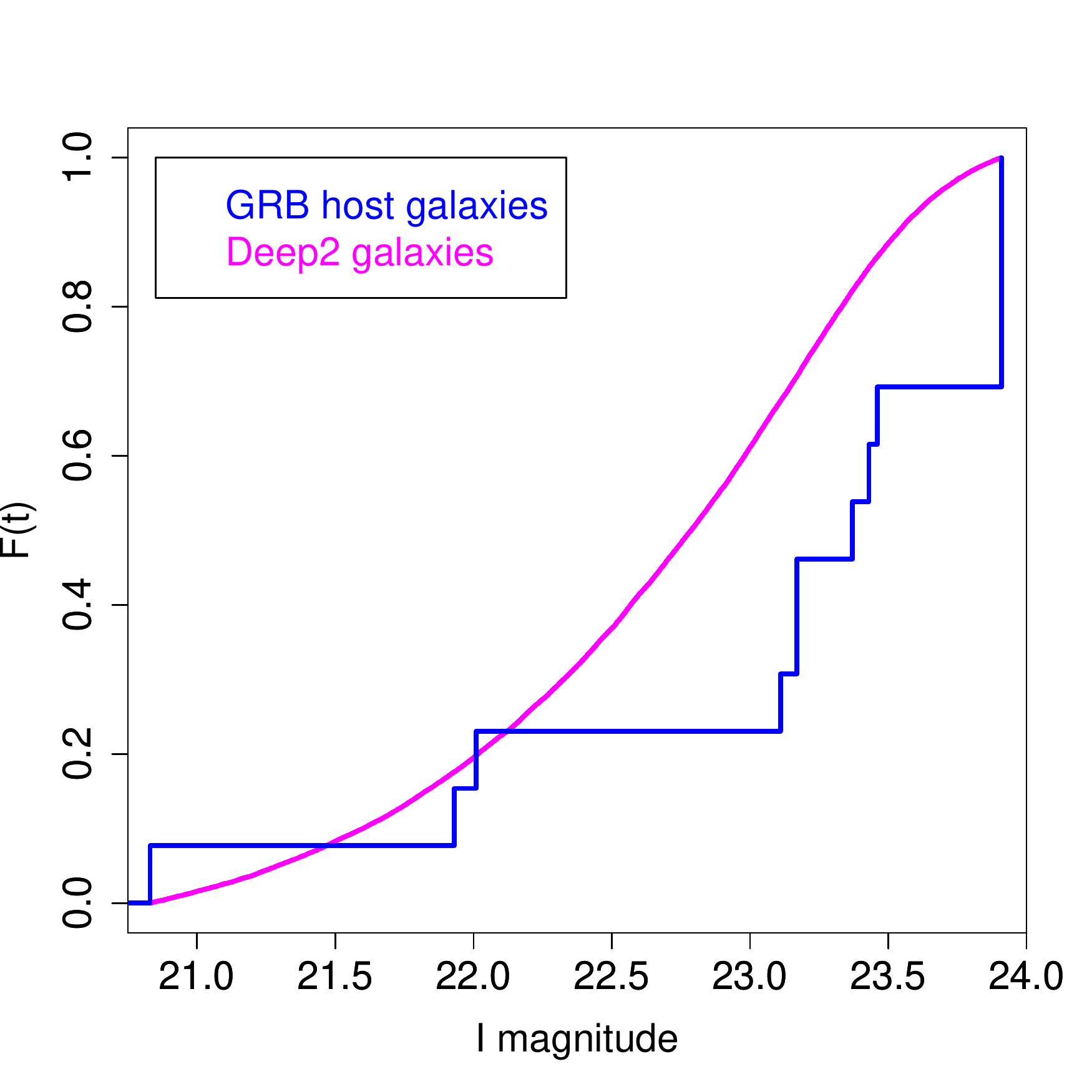}
\includegraphics[width=.51\textwidth]{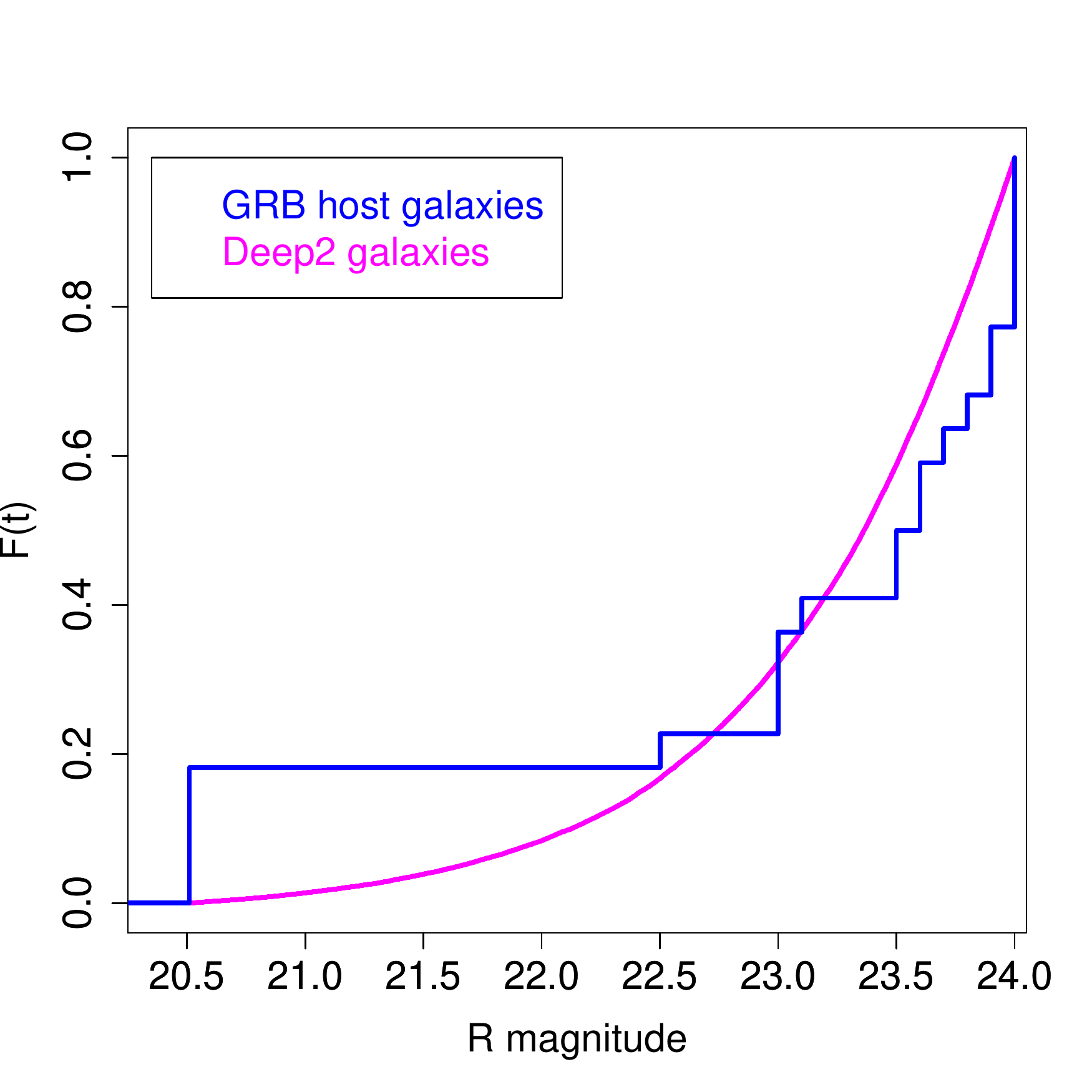}
\caption{The comparison of the distribution function obtained from the survival function I color and the DEEP2 I color distribution (left) and R color (right). The brightness distribution of the hosts are systematically fainter than the DEEP2 galaxies, the probability of random difference is $\mathrm{9\cdot10^{-5}}$ in R band and only $\mathrm{3\cdot10^{-5}}$ in I band.}
\label{fig3}
\end{figure}

Since the GRBs are connected to the standard candle luminosity function assumption  \cite{horvath1, horvath2}, we compared the HGs luminosity distribution with DEEP2 galaxies sample between about 21 and 24 magnitude on both colors. Significant difference in the two cases were found. The probabilities of random differences are only $\mathrm{3\cdot10^{-5}}$ in I band and $\mathrm{9\cdot10^{-5}}$ in R band. The results are shown in Figure~\ref{fig3}. We found that the probability is three times smaller in I color meaning that the GRB host galaxies are bluer than normal galaxies.

Table \ref{tab1} shows the result from the basic Cox regression. We didn't find significant correlations between the GRB host brightness and some GRB physical parameters as the BAT T90, Fluence and Photon index of the GRBs. But we didn't yet examine all off physical parameters of GRBs we would like to end this work.

When we tested the CPL-PL groups difference in host galaxies with log-rank or Cochran-Mantel-Haenszel test we found there are no differences between the two classes, the probability being 0.73 in R and 0.5 in I band.

Finally, almost the same results were found when we examined that if the redshifts are known, the probabilities are around 0.5. This result is very surprising because it has been expected that the redshifts of the fainter host galaxies would be less known.

\begin{table}[!h]
\centering
\begin{tabular}{|l|rc|rc|}
\hline 
Variable & \multicolumn{2}{c|}{I color} & \multicolumn{2}{c|}{R color} \\  
 & $\mathrm{\beta_{coeff}}$ & Prob & $\mathrm{\beta_{coeff}}$ & Prob \\ 
\hline 
$\log(1+z)$ & $-$1.125 & 0.46 & $-$1.406& 0.21 \\ 
\hline 
$\log(T90)$ & 0.182 & 0.83 & $-$0.129 & 0.85 \\ 
\hline 
$\log(Fluence)$ & $-$1.685 & 0.17 & $-$0.540 & 0.62 \\ 
\hline 
$\log(Peak Flux)$ & 0.462 & 0.68 & 0.424 & 0.63 \\ 
\hline 
$\log(Photon index)$ & $-$2.675 & 0.36 & 1.923 & 0.47 \\ 
\hline 
$\log(XRT_{11h})$ & $-$1.621 & 0.46 & 1.072 & 0.55 \\ 
\hline 
$\log(XRT_{24h})$ &1.578 & 0.42 & $-$0.838 & 0.60 \\ 
\hline
\end{tabular} 
\caption{Results of the Cox regressions. The Prob columns gives the probability, the dependence values are fully random. Seemingly, no remarkable correlations were obtained.}
\label{tab1}
\end{table}


\section{Conclusion and Summary}
Comparing the cumulative distribution functions resulted in the survival analysis of the GRB hosts  with that obtained from the DEEP2 sample in the R and I colors we recognized an offset of hosts  towards fainter magnitudes. This offset is significant at the $\mathrm{9\cdot10^{-5}}$ level in the R and $\mathrm{3\cdot10^{-5}}$ in the I color. Assuming the same spatial distribution of the GRB hosts and the DEEP2 sample galaxies this offset is explained by the lower absolute brightness and probably systematically lower mass of the GRB hosts. Moreover, the offset in the R seems to be smaller than in the I color indicating a bluer R-I color index of the GRB hosts. The bluer color index  may indicate a higher star forming activity in the hosts then in the galaxies of the DEEP2 sample.

Moreover, we studied dependence of the GRB host galaxies brightness distribution on the GRB SWIFT data as SWIFT BAT T90, Fluence, 1-sec Peak Photon Flux, Photon Index, XRT 11 Hour Flux, XRT 24 Hour Flux and redshift and it wasn't found significant relationship between the data. Finally, no significant differences were found between HGs with PL or CPL GRB spectra or the redshift of host galaxies (redshift of GRBs) are known or not.

\acknowledgments
This work was supported by the Hungarian OTKA NN-111016 grant.

\end{document}